\documentclass{iau}
\usepackage{graphicx}
\usepackage{bm}

\title[Flux emergence event underneath a filament] 
{Flux emergence event underneath a filament}

\author[Palacios et. al.]   
{J. Palacios$^1$, Y. Cerrato$^1$, C. Cid$^1$, A. Guerrero$^1$, and 
 E. Saiz$^1$}

\affiliation{$^1$Space Reseach Group--Space Weather, Departamento de F\'isica y Matem\'aticas, \\ Universidad de Alcal\'a \\ 
University Campus, Sciences Building, P.O. 28871, Alcal\'a de Henares, Spain \\
 email: {\tt judith.palacios@uah.es}}

\pubyear{2015}
\volume{305}  
\jname{Polarimetry: From the Sun to Stars and Stellar Environments}
\editors{K.N. Nagendra, S. Bagnulo, \\ R. Centeno, \& M. Mart\'inez Gonz\'alez, eds.}
\begin{document}

\maketitle

\def\arcsec{\hbox{$^{\prime\prime}$}}
\def\deg{\mbox{$^{\circ}$}}

\begin{abstract}

Flux emergence phenomena are relevant at different temporal and spatial scales. We have studied a flux emergence region underneath a filament. This filament elevated itself smoothly, and the associated CME reached Earth. In this study we investigate the size and amount of flux in the emergence event. The flux emergence site appeared just beneath a filament. The emergence acquired a size of 24 Mm in half a day. The unsigned magnetic flux density from LOS-magnetograms is around 1 kG at its maximum. The transverse field as well as the filament eruption were also analysed. 

\keywords{Sun: photosphere, magnetic fields, filaments; techniques: polarimetric}
\end{abstract}

\firstsection 

\section{Introduction}
\label{intro}

Flux emergence is one of the most important events in the solar photosphere, acting as part of the local dynamo or small-scale surface dynamo (e.g. \cite[Mart{\'{\i}}nez Pillet, 2013]{Valentin2013-pal1}). Sometimes the ubiquitous circumstance of flux emergence can be of such a large scale that may affect structures in the chromosphere and low corona. For an extensive review on the topic, see \cite[Cheung \& Isobe (2014)]{Cheung2014-pal1}.The different spatial and temporal scales of flux emergence features have been reviewed by \cite[Schmieder et al. (2014)]{Schmieder2014-pal1}.

The characteristics of emerging magnetic flux --  basically its area and its magnetic flux density -- and the total time employed by the feature to emerge, can be combined to get a temporal flux emergence growth rate for these events, as will be mentioned in the following paragraphs. Magnetic flux is measured in Weber (Wb), but Maxwell (Mx) is commonly used. The equivalence is 10$^{8}$ Mx to 1 Wb. Its temporal rate is given in Mx~day$^{-1}$ (or other unit of time) here.

Flux emergence episodes have been investigated in detail in the era of high-resolution spectropolarimetry. The largest cases, as sunpot emergence, have been studied thoroughly, but smaller ones have remained evasive until recent times, due to the low spatial and spectral resolution of the observations. At the smallest scales, there are some studies at the granular level, as in \cite[De Pontieu (2002)]{DePontieu2002-pal1}, \cite[Orozco et al. (2008)]{Orozco2008-pal1}. Granular-scale simulations have also been performed, e.g., in \cite[Tortosa \& Moreno-Insertis (2009)]{Tortosa2009-pal1}. At the mesogranular scale, flux emergence cases found by \cite[Palacios et al. (2012)]{Palacios2012-pal1} shows growth rates of 10$^{18}$ Mx in about 20 min. Spectral line inversions show field strengths of hectogauss in these mesogranular-sized patches (see \cite[Guglielmino et al. (2013)]{Guglielmino2013-pal1}. The flux emergence rates at these spatial scales from longitudinal synthetic magnetograms are in the same range, $\sim$~10$^{17}$~Mx~min$^{-1}$ (\cite[Cheung et al. 2008]{Cheung2008-pal1}). 

Some of these small emergence cases come in the form of loop-like features: two opposite polarities linked by a patch of transverse field. An amount of cases have been described in, e.g., \cite[Centeno et al. (2007)]{Centeno2007-pal1}, \cite[G\"om\"ory et al. (2010)]{Gomory2010-pal1}, \cite[Palacios et al. (2012)]{Palacios2012-pal1}. Loop statistics, lifetimes and energy implications are described in \cite[{Mart{\'{\i}}nez Gonz{\'a}lez} et al. (2009)]{Marian2009-pal1} and \cite[{Mart{\'{\i}}nez Gonz{\'a}lez} et al. (2010)]{Marian2010-pal1}. In these very small emergence cases, the flux emergence rate is $\sim$ 10$^{17}$ Mx~min$^{-1}$.

We should also mention the ephemeral regions (EFR) which are manifestations of small bipolar emergence that constitute an important contribution to the solar surface magnetism. The typical flux emergence rate per feature is $\sim$ 10$^{15}$ Mx~s$^{-1}$, with total flux contents in the smaller EFRs of 10$^{18}$ Mx. These features typically emerge over the course of several hours (\cite[Guglielmino et al. 2012]{Guglielmino2012-pal1}), and contribute with 10$^{22}$ Mx~day$^{-1}$ to the total flux on the solar surface (\cite[Hagenaar, 2001]{Hagenaar2001}). These regions are also taken into account for the calculation of the total solar irradiance, as shown in \cite[Krivova et al. (2007)]{Krivova2007-pal1}, among others. Other examples of small bipolar emergence fields are the  serpentine fields -- undulatory fields --, exhibiting U- and $\Omega$-shaped loops. \cite[Vargas Dom{\'{\i}}nguez et al. (2012)]{Santi2012-pal1} showed some observational examples. \cite[Valori et al. (2012)]{Valori2012-pal1} made force-free extrapolations of sea-serpent magnetic fields in the context of a large active region emergence.

Very recent large-scale flux emergence events have been investigated. Being intermediate to large sized active regions, the flux emergence growth rate may vary but it is around 10$^{21}$ Mx~day$^{-1}$ for active region emergence simulations, as in \cite[Rempel \& Cheung (2014)]{Rempel2014-pal1}, and for observations, as in \cite[Centeno (2012)]{Centeno2012-pal1}. One of the largest simulated ARs can be found in \cite[Cheung et al. (2010)]{Cheung2010-pal1}, being almost an order of magnitude larger in flux emergence rate. Other interesting cases of active region emergence are the `anemones' or fountain-like regions (\cite[Asai et al. 2008]{Asai2008-pal1}). These events are related to active regions growing into coronal hole field, which consist of mostly open field lines extending into the interplanetary medium and acting as an ambient field that leads to reconnection. 

Flux rope emergence is one of the occurrences with larger consequences for space-weather, since it may further trigger eruptions and CMEs in the aftermath. This emergence may be caused by whole body emergence or further evolution of the magnetic field in an active region (see \cite[Cheung \& Isobe 2014]{Cheung2014-pal1}). Eventually it may provoke an eruption. These enormous eruptions may be caused by different physical mechanisms. Some of these processes do include flux emergence and cancellation by reconnection. For an exhaustive review, see e.g. \cite[Zuccarello et al. (2013)]{Zuccarello2013-pal1}. There are some reported cases of filament eruption and flux emergence (e.g. \cite[Wang \& Sheeley 1999]{Wang1999-pal1}); still, the direct cause-effect relationship between flux emergence and filament eruption is not that clear. 
\cite[Cid et al. (2014)]{Cid2014-pal1} and references therein address the role of these eruptive phenomena and the geoeffectiveness of these features in extreme geomagnetic storms.

In this paper we focus on the case of the emergence of photospheric magnetic fields (using spectropolarimetric data) along with some detail on the eruption itself.

\section{Data}

We have mainly used data from the Helioseismic and Magnetic Imager (HMI, \cite[Scherrer et al. 2008]{Scherrer2008-pal1}) on board the Solar Dynamics Observatory (SDO, \cite[Pesnell et al. 2012]{Pesnell2012-pal1}) in the form of HMI magnetograms and full Stokes images, from 2013, 29 Sept. This analysis was also completed with data from AIA/SDO (\cite[Lemen et al. 2012] {Lemen2012-pal1}) 171, 1600~\AA. Data from the LASCO C2 coronagraph (\cite[Brueckner et al. 1995]{Brueckner1995-pal1}) onboard SOHO were checked for the eruption as well.

\section{Flux emergence}

\begin{figure}[t]
\begin{center}
\includegraphics[scale=0.65]{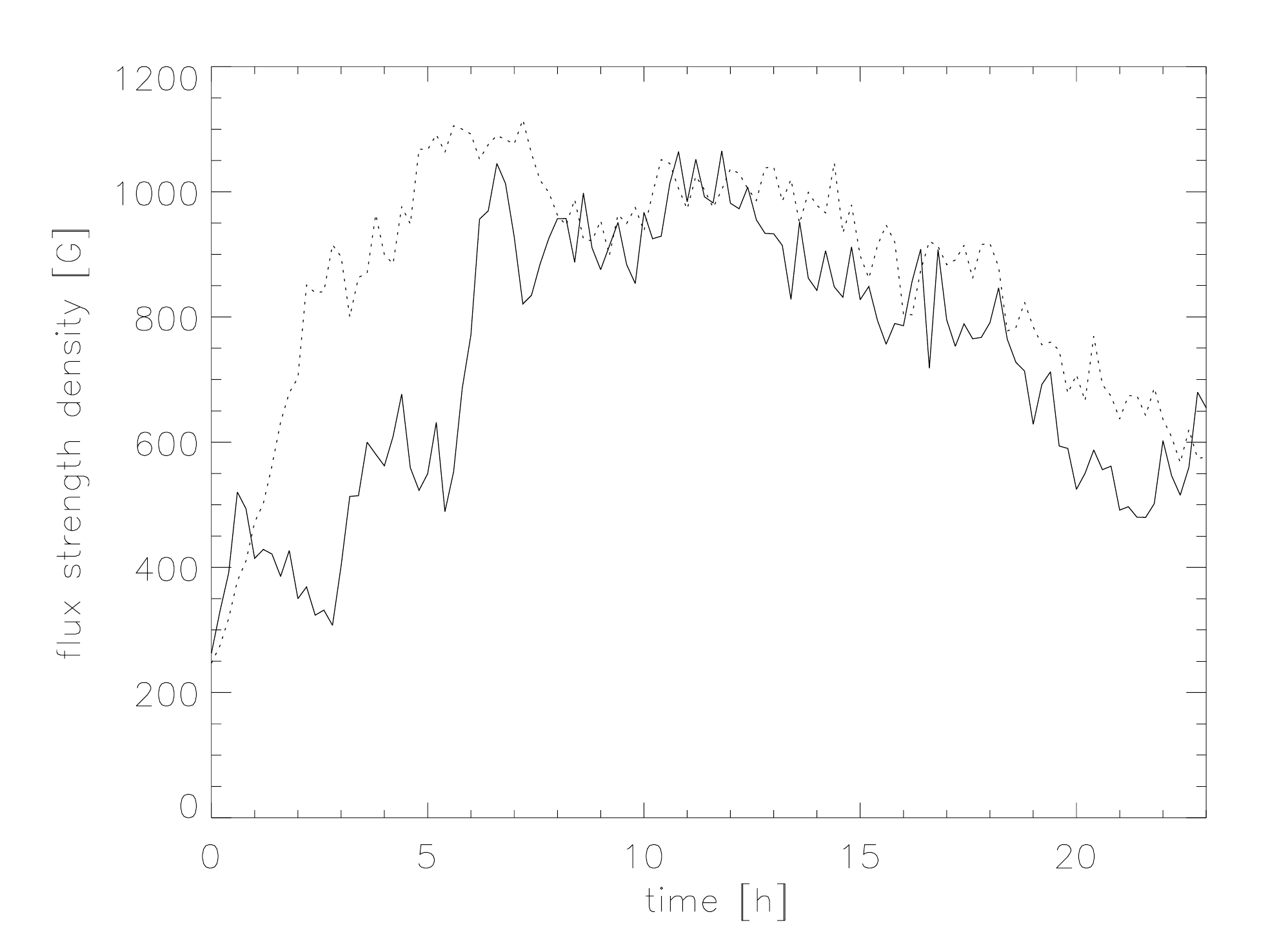}
\caption{Maximum magnetic flux density from 00:00 UT to 23:00 UT. The solid line indicates the maximum (positive) magnetic flux density and the dashed line, the minimum (negative) magnetic flux density in absolute value.
}
\label{bhmi-pal1}
\end{center}
\end{figure}

We have observed the emergence event on HMI data underneath a filament using HMI data. We analysed the magnetic flux density from the line-of-sight (LOS) magnetograms. Fig.~\ref{bhmi-pal1} shows the maximum and minimum of this quantity, from 00:00 UT to 23:00 UT on Sept 29. These values are corrected for heliocentric angle projection ($\mu$). Some network sub-structures exhibit longitudinal magnetic flux densities of the order of kilogauss (kG) during several hours.

\begin{figure}[t]
\begin{center}
\includegraphics[scale=0.31]{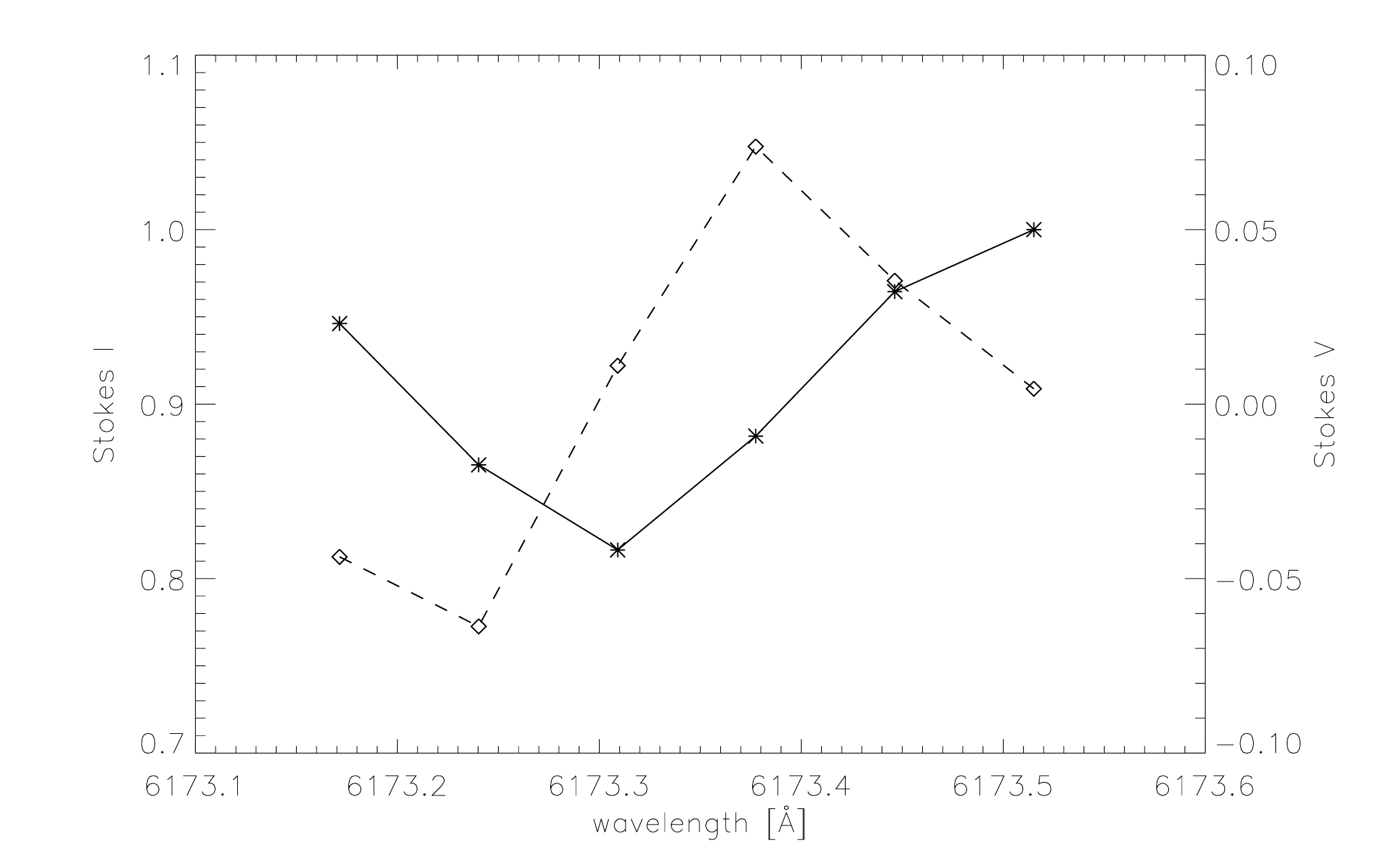}
\includegraphics[scale=0.31]{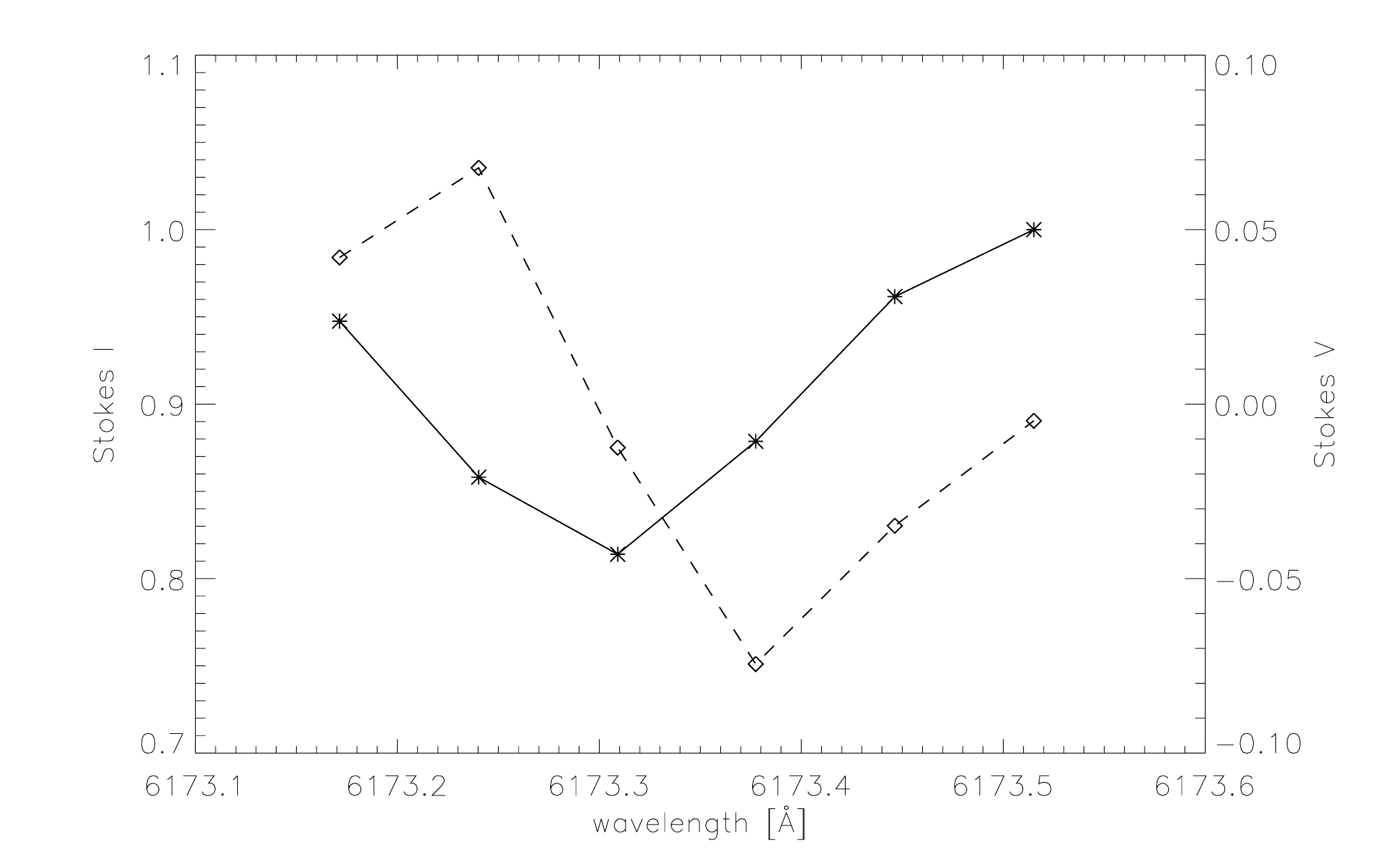}

\caption{Stokes {\it I} (solid) and {\it V} (dashed, Y axis on the right) from the pixels that exhibit the largest extreme values in magnetic field strength ({\it left}: negative, {\it right}: positive).}
\label{stks-pal1}
\end{center}
\end{figure}

Taking advantage of the full Stokes \textit{[I, Q, U, V]} imaging from HMI, we analysed the linear and circular polarization. We normalized the spectral profiles coarsely using the redmost wavelength point of Stokes {\it I}, and used the wavelength centering of this line (6173.34~\AA, spacing 69~m\AA, \cite[Hoeksema et al. 2014]{Hoeksema2014-pal1}). Fig.~\ref{stks-pal1} shows the Stokes {\it I} (solid) and  Stokes {\it V} (dashed) profiles corresponding to the pixel that display the largest negative and positive values in the magnetogram (note that the Stokes {\it V} profiles are clearly antisymmetric). To obtain the magnetic field strength density, we have employed the weak-field approximation (\cite[Degl'Innocenti \& Landolfi 2004]{Innocenti2005-pal1}) with the HMI Stokes profiles. This leads also to kilogauss values in some pixels, in agreement to the HMI LOS-magnetograms. 

Most likely, a mechanism of magnetic field intensification in the form of `convective collapse' (\cite[Parker 1978]{Parker1978},  \cite[Spruit 1979]{Spruit1979-pal1}) took place in this network patch, since it exhibits signals of small pore growth processes. For some case studies on this physical mechanism, see e.g., \cite[Vargas Dom{\'{\i}}nguez et al. (2015)]{Santi2015-pal1} and references therein.

\begin{figure}[t]
\begin{center}
\includegraphics[scale=0.50]{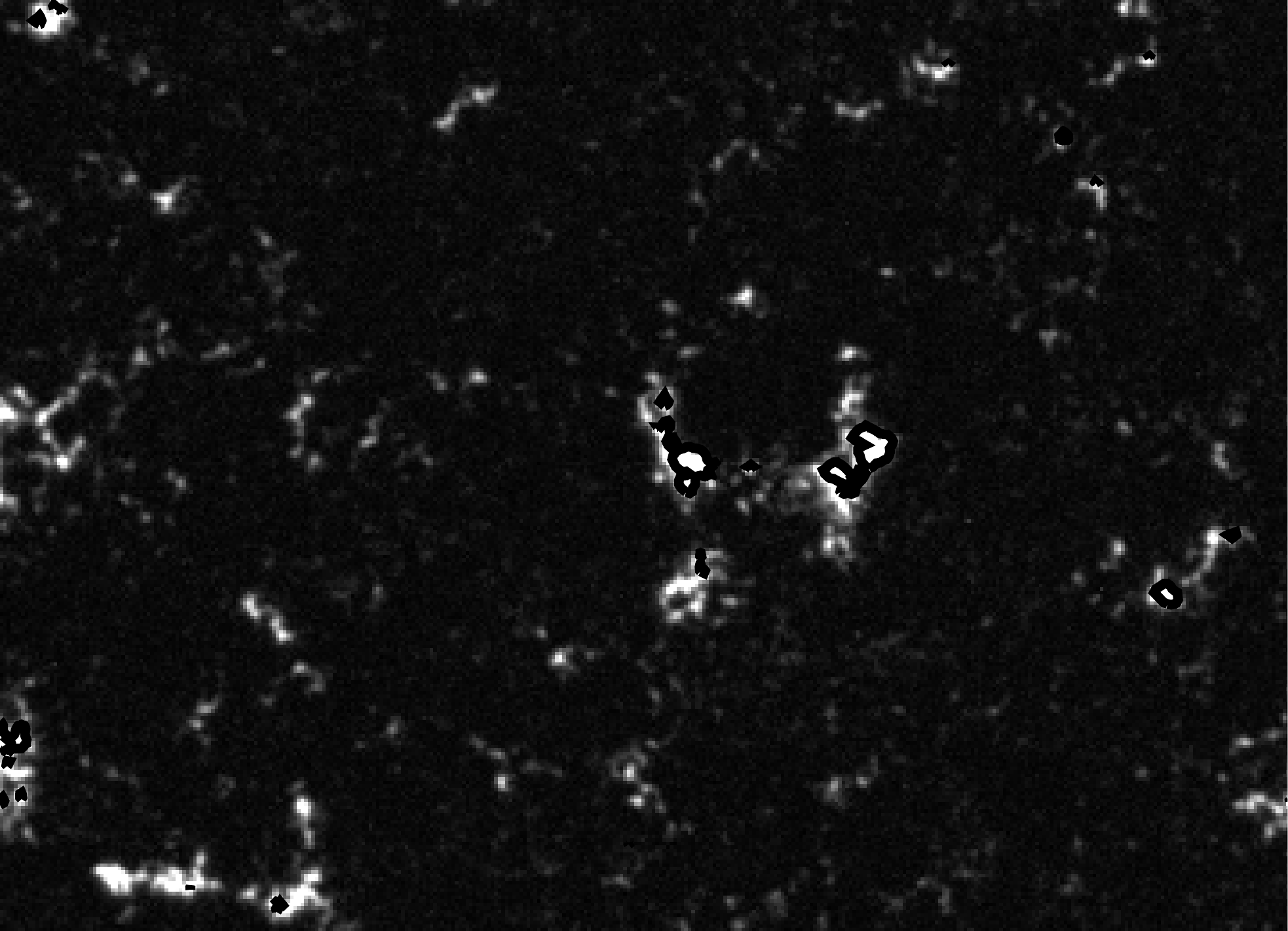}
\caption{Background image: Absolute value of the maximum of Stokes {\it V} normalized to the continuum intensity. The image is clipped to 3\% for clarity. The thick line contour represents the linear polarization at 0.25\% level. }
\label{mappol-pal1}
\end{center}
\end{figure}

We also computed the linear and circular polarization from these Stokes profiles. Fig.~\ref{mappol-pal1} displays an image of the circular polarization at 10:00 UT, with some linear polarization contours overplotted. The background image corresponds to the maximum (in absolute value) of Stokes {\it V} (normalized to the continuum intensity), clipped at 0.03; the linear polarization contours mark the 0.0025 level with respect to the continuum intensity. The field of view (FOV) is $\sim$ 150\arcsec $\times$ 110\arcsec.

\section{Eruption}

\begin{figure}[t]
\begin{center}
\includegraphics[scale=0.32]{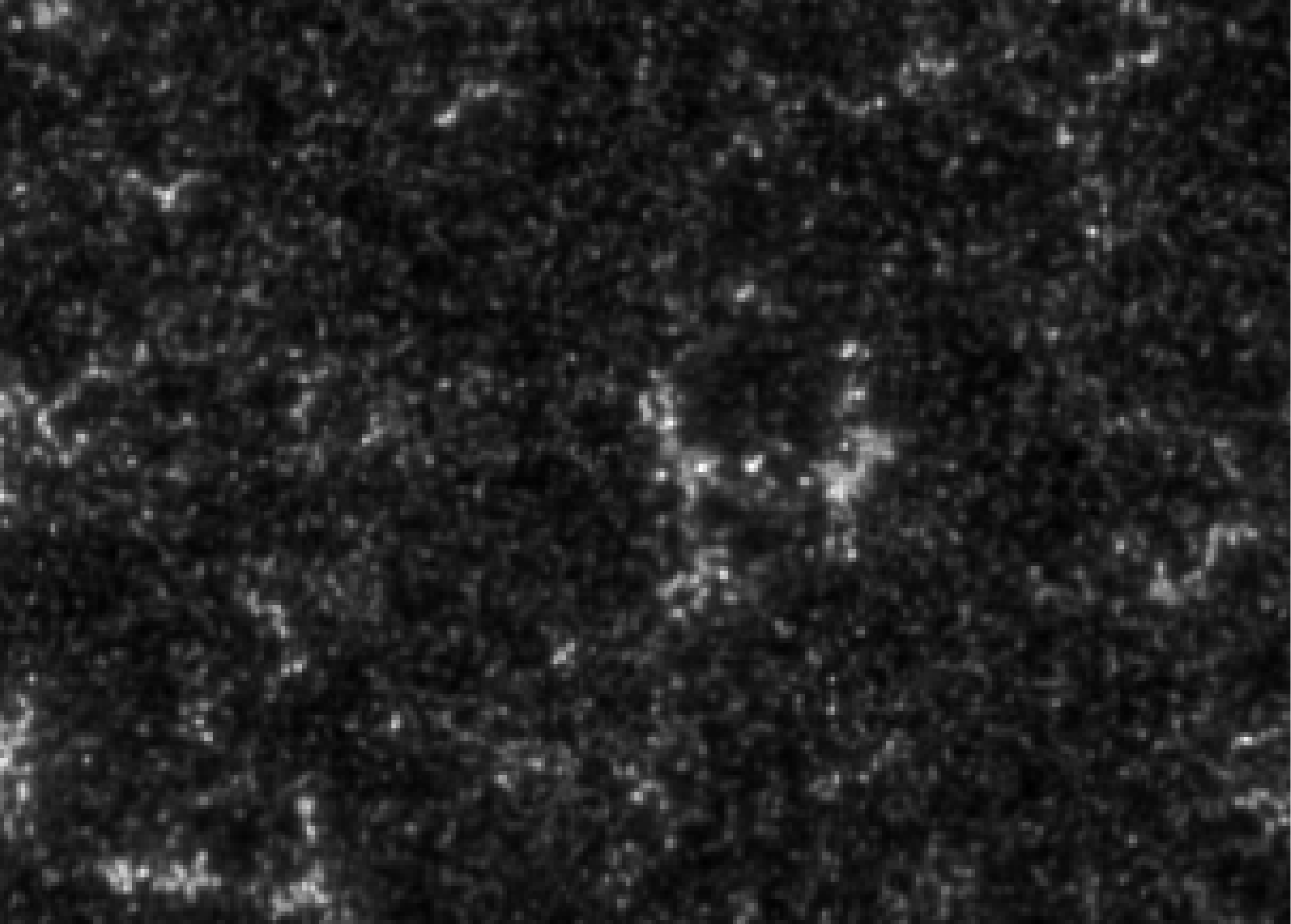}
\includegraphics[scale=0.32]{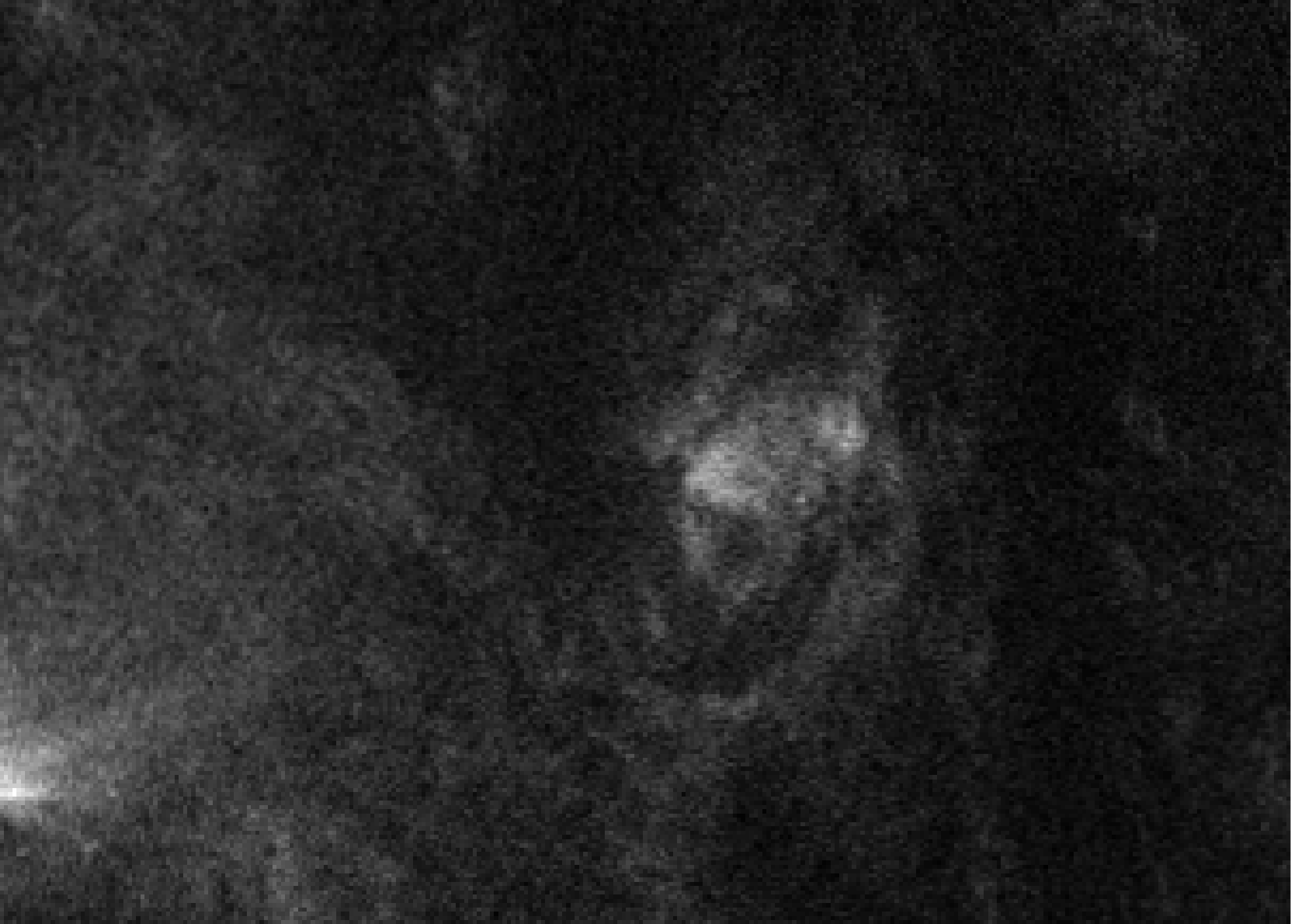}
\caption{{\it Left:} Image in 1600~\AA~at 10:00~UT. {\it Right:} Image in 171~\AA~in logarithmic scale at the same moment.}
\label{aiaimag-pal1}
\end{center}
\end{figure}

We also studied the evolution of different solar layers in order to know the correspondence of this flux emergence at different heights. We used AIA images from two different wavelengths, 1600 and 171 \AA, which correspond to the high photosphere and corona, respectively. These are shown in Fig.~\ref{aiaimag-pal1}. The field of view (FOV) is also $\sim$ 150\arcsec $\times$ 110\arcsec.

The image in 1600~\AA~(Fig.~\ref{aiaimag-pal1}, left) shows a large network patch, the same as the one shown in Stokes {\it V} image of Fig.~\ref{mappol-pal1}, surrounded by other network areas. However, images in logarithmic scale in 171~\AA~clearly display a brighter area, dipolar-shaped. (Fig.~\ref{aiaimag-pal1}, right). The filament is not seen in 171~\AA, since this channel is best for viewing the quiet corona because it has a contribution for higher temperatures (\cite[O'Dwyer et al. 2010]{ODwyer2010-pal1}). The eruptive filament is clearly seen in high chromosphere - transition region wavelengths.
We also looked for signatures of flux rope photospheric emergence on data a few days before the eruption, following e.g. \cite[Kuckein et al. (2012)]{Kuckein2012-pal1}, but unfortunately we could not detect it.

\begin{figure}[t]
\begin{center}
\includegraphics[scale=0.45]{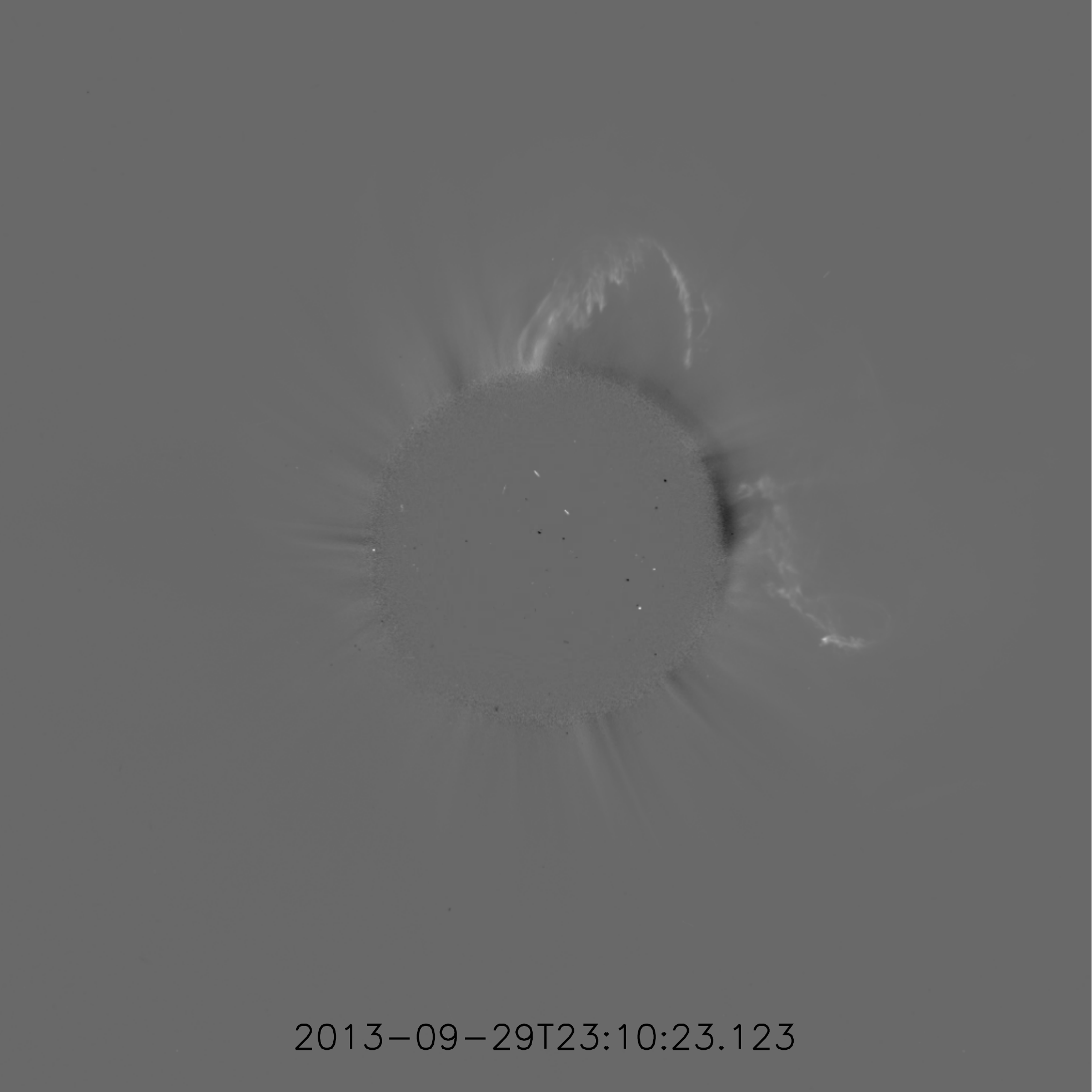}
\caption{Image of the filament as observed by LASCO C2.}
\label{lascoc-pal1}
\end{center}
\end{figure}

To have a wider FOV around the eruption, LASCO data is also employed. LASCO C2 data field of view ranges from 2 to 6 solar radii. Fig.~\ref{lascoc-pal1} shows a LASCO C2 image of the eruption of this filament more than one hour after the eruption. It displays a twisted flux-rope structure in the solar north-west. The highest point of the filament in this image is $\sim$ 4 solar radii over the surface. The eruption speed of the filament is on the order of hundreds of km s$^{-1}$.

\section{Final remarks}

In this paper we have a flux emergence event related to a filament eruption, studying longitudinal flux density, and the linear and circular polarization with HMI/SDO data, in addition to height and shape on LASCO/SOHO imaging. We have also investigated the emergence in different wavelengths of AIA, namely, 171 and 1600 \AA, which show excellent correspondence in the high photosphere and corona. A detailed study of this event will appear in a forthcoming paper.

 \section{Acknowledgements}
We are very grateful to HMI, AIA and LASCO teams for all valuable data used in this work. J.~P.~acknowledges funding from IAU to attend IAUS305 and UAH-travel grants. She also acknowledges project AYA2013-47735-P.

\end{document}